\documentclass[twocolumn,aps,showpacs,prb]{revtex4}
\usepackage{graphicx}
\usepackage{epstopdf}
\usepackage{amsmath}
\usepackage{multirow}

\begin{document}

\title{Atomic and electronic structures of FeSe monolayer and bilayer
thin films on SrTiO$_3$ (001): a first-principles study}


\author{Kai Liu$^{1}$}
\author{Zhong-Yi Lu$^{1}$}
\author{Tao Xiang$^{2,3}$}

\date{\today}

\affiliation{$^{1}$Department of Physics, Renmin University of China, Beijing 100872,
China}

\affiliation{$^{2}$Institute of Theoretical Physics, Chinese Academy of Sciences, Beijing
100190, China }

\affiliation{$^{3}$Institute of Physics, Chinese Academy of Sciences, Beijing 100190,
China }

\begin{abstract}

By the first-principles electronic structure calculations, we have
studied electronic structures of FeSe monolayer and bilayer thin
films on SrTiO$_3$ (001) with SrO termination or TiO$_2$
termination. We find that both FeSe monolayer and bilayer on either
termination behave like a slightly doped semiconductor and a
collinear antiferromagnetic order on Fe ions. There is no
substantial charge transfer between the FeSe layers and the
substrate. FeSe is adhered to the SrTiO$_3$ surface by a
dipole-dipole interaction. The Fermi surface is mainly the
contribution of Fe-$3d$ orbitals. A valence band contributed mainly
by the O-$2p$ orbitals in the TiO$_2$ layer is located slightly
below the Fermi level, which can become conducting upon a small
doping of holes.

\end{abstract}

\pacs{68.35.B-, 73.20.-r, 74.70.Dd, 68.43.Bc}

\maketitle

\section{INTRODUCTION}

The discovery of Fe-based superconductors\cite{1111,122,111,11} has
inspired worldwide interests both experimentally and theoretically
in the recent years. The common feature of all the Fe-based
superconductor compounds is that they have FeX layers (X=P, As, S,
Se or Te) composed of edge-sharing tetrahedra with an Fe center.
Furthermore, some of the parent compounds of these superconductors
are antiferromagnetic (AFM) semimetals \cite{lu1} with either a
collinear \cite{cruz,ma1} or bi-collinear \cite{ma,bao} AFM order
below a structural transition temperature. The superconductivity can
be induced either by hole or electron doping via chemical
substitution or by high pressure, and the superconducting paring is
believed to occur in the FeX layers.

In Fe-based superconductors, it was widely believed that the
As-bridged antiferromagnetic superexchange interaction between the
next nearest neighbors Fe-Fe atoms plays an important
role\cite{ma1,seo}. This kind of interaction depends strongly on the
local geometry of a Fe-As bond. Indeed, it was found that the
geometry of the FeX$_4$ tetrahedral unit, in particular the bond
length and the bond angle between two neighboring X and Fe ions,
strongly correlates with the superconductivity transition
temperature\cite{yamada,cava}. Besides applying pressure, growing an
Fe-based superconductor on some substrate is another effective way
to manipulate the lattice parameters and the FeX$_4$ tetrahedral
geometry, which may tune the superconducting transition temperature.
As an orthodox metal oxide perovskite, Strontium titanate
(SrTiO$_3$) is widely used as a high-quality substrate for epitaxial
growth of high-temperature cuprate superconductors and many other
oxide-based thin films\cite{ueno}. Very recently, SrTiO$_3$ has been
used as a substrate to epitaxially grow FeSe ultrathin films
\cite{xue}. It is found that the monolayer FeSe thin film grown on
SrTiO$_3$ (001) shows signatures of superconducting transition above
50 K by transport measurement, while the bilayer and thicker films,
in contrast, do not show any sign of superconductivity\cite{xue}.

To understand the physics underlying this remarkable finding, we
have performed the first-principles electronic structure
calculations on FeSe monolayer and bilayer ultrathin films on
SrTiO$_3$ (001), respectively. We find that both monolayer and
bilayer thin films show a semiconducting behavior with a collinear
AFM order on Fe atoms, substantially different from the metallic
bulk FeSe. Considering the difference between the single FeSe and
multi-layer FeSe thin films, this suggests that the observed
superconductivity happens either at the interface of FeSe-SrTiO$_3$
or just in the first FeSe layer, not in the other FeSe layers.

Crystal SrTiO$_3$ is built from alternatively stacking planar SrO
and TiO$_2$ square layers along the $c$-axis. Experimentally, bulk
SrTiO$_3$ takes a structural phase transition from cubic perovskite
structure into a tetragonal one at 105K, in which each octahedral
unit TiO$_6$ with titanium centered slightly rotates around the $z$
axis, meanwhile the rotation directions between the nearest
neighboring octahedral units are reverse with each other
\cite{unoki,palmer99}. Such a structural distortion is thus called
antiferrodistortion.

The cleaving of SrTiO$_3$ simultaneously generates both
SrO-terminated and TiO$_2$-terminated surfaces with the
corresponding cleavage energy equally distributed between these two
surfaces \cite{muller,borstel1}. It was further shown that the
surface energies are nearly the same for these two surfaces after
full relaxation \cite{borstel1,borstel2}. This means that both
surfaces are stable and either of them may be used as a substrate to
grow FeSe thin films. We thus study FeSe monolayer and bilayer on
both TiO$_2$-terminated and SrO-terminated SrTiO$_3$ (001),
respectively.

\section{COMPUTATIONAL DETAILS}

To study the atomic structures and the electronic and magnetic
properties of FeSe ultrathin films on SrTiO$_3$ (001), we carried
out the fully spin-polarized first-principles electronic structure
calculations by using the projector augmented wave (PAW) method
\cite{kresse,paw}. We adopted the generalized gradient approximation
of Perdew-Burke-Ernzerhof \cite{pbe} for the exchange-correlation
potentials. After the full convergence test, the kinetic energy
cut-off of the plane wave basis was chosen to be 400 eV. The
optimization made the forces on all relaxed atoms smaller than 0.02
eV/\AA .

We first checked the properties of bulk SrTiO$_3$. We constructed a
20-atom tetragonal supercell so that we can describe the
antiferrodistortive structure, which unit cell is $\sqrt{2}\times
\sqrt{2}\times 2$ unit cell of undistortive bulk SrTiO$_3$. The
Brillouin zone was sampled with a $6\times 6\times 6$ {\bf k}-point
mesh. After both the shape and volume of the supercell and the
internal atomic positions were optimized, we find that SrTiO$_3$ has
indeed an antiferrodistortive tetragonal structure with the lattice
parameters $a^{\star}=b^{\star}=\sqrt{2}a=5.536~$\AA~ and
$c^{\star}=2c=7.831~$\AA~ ($a$ and $c$ being the lattice parameters
of the undistortive tetragonal unit cell), the octahedral rotation
angle $\theta$=1.1$^{\circ}$, and the ratio
(c/a)-1=3$\times$10$^{-4}$. These values agree well with the
experimental data \cite{palmer99} $a^{\star}=b^{\star}=5.507~$\AA,
$c^{\star}=7.796~$\AA, $\theta$=2.1$^{\circ}$, and $(c/a)-1=10\times
10^{-4}$, respectively, as well as with the previous calculation
results \cite{scuseria11}.

To model FeSe ultrathin films on TiO$_2$ (SrO)- terminated SrTiO$_3$
(001), we used a 6 (7)-layer SrTiO$_3$(001) slab with FeSe monolayer
and bilayer FeSe adsorbed on the top side in a
$\sqrt{2}a\times\sqrt{2}a$ two-dimensional supercell plus a vacuum
layer of about 10 \AA. The top two slab layers and all FeSe layer
atoms were allowed to relax while the rest slab layers were fixed at
their bulk positions. A $6\times 6\times 1$ {\bf k}-point mesh for
the Brillouin zone sampling and the Gaussian smearing technique with
a width of 0.1 eV were used. The electric field induced by
asymmetric slab relaxation was compensated by a dipole correction
\cite{scheffler92}.

\section{RESULTS AND ANALYSIS}

\begin{figure}
\includegraphics[angle=0,scale=0.25]{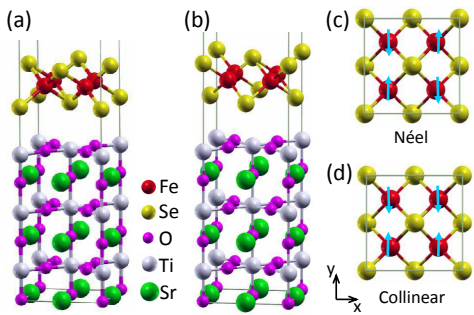}
\caption{(Color online) Atomic structures (a)I and (b)II of
monolayer FeSe on TiO$_2$-terminated SrTiO$_3$(001) surface.
Patterns of Fe atom spins in the (c)checkerboard antiferromagnetic
N\'eel state and (d)the collinear antiferromagnetic state.
}\label{fig1a}
\end{figure}

\begin{figure}[!b]
\includegraphics[angle=0,scale=0.25]{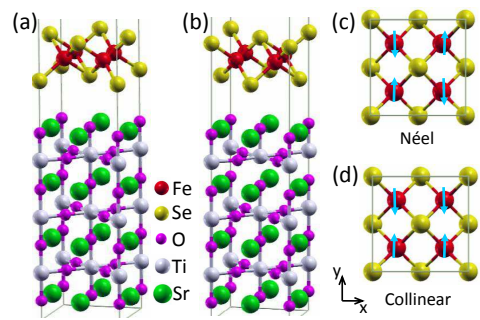}
\caption{(Color online) Atomic structures (a)I and (b)II of
monolayer FeSe on SrO-terminated SrTiO$_3$(001) surface. Patterns of
Fe atom spins in the (c)checkerboard antiferromagnetic N\'eel state
and (d)the collinear antiferromagnetic state.}\label{fig2a}
\end{figure}

For both TiO$_2$-terminated and SrO-terminated SrTiO$_3$ substrates,
we studied two possible adsorption structures of FeSe thin films as
shown in Fig. \ref{fig1a} (a) and (b) and Fig. \ref{fig2a} (a) and
(b), respectively. The nonmagnetic, the ferromagnetic, the
checkerboard antiferromagnetic N\'eel, and the collinear
antiferromagnetic states were all calculated. Patterns of Fe atom
spins in the checkerboard antiferromagnetic N\'eel and the collinear
antiferromagnetic states are shown in panels (c) and (d) of Fig.
\ref{fig1a} and Fig. \ref{fig2a}, respectively. To better understand
the electronic properties of FeSe thin films on SrTiO$_3$, the
electronic band structures of clean SrTiO$_3$ surface and isolated
FeSe monolayer are firstly introduced in subsection A. Results of
monolayer and bilayer FeSe ultrathin films on TiO$_2$-terminated and
SrO-terminated SrTiO$_3$ are then presented in subsection B and C,
respectively.

Figure \ref{fig1a} shows the atomic structures of monolayer FeSe on
TiO$_2$-terminated SrTiO$_3$ surface. For structure I in panel (a),
Se atom locates on top of the Ti atom, which substitutes the
position of O atom in TiO$_6$ octahedron of bulk SrTiO$_3$. For
structure II in panel (b), Se atom sits on the center site of the
Ti-Ti square. The atomic structures of monolayer FeSe on
SrO-terminated SrTiO$_3$ surface are shown in Fig. \ref{fig2a}. For
structure I in panel (a), Se atom locates on top of the O atom. As
to the structure II in panel (b), Se atom is on top of the Sr atom.
By considering the symmetry of both the ad-layer and the substrate,
these four structures in Figs. \ref{fig1a} and \ref{fig2a} are the
most possible adsorption sites.

\subsection{Clean SrTiO$_3$ surface and isolated FeSe monolayer}

In order to have a concept about how the electronic structures of
FeSe-SrTiO$_3$ heterostructures are derived from their subdivisions,
we firstly studied the band structures of TiO$_2$- and
SrO-terminated SrTiO$_3$(001) clean surfaces and isolated FeSe
monolayer. For SrO-terminated surface, the atomic structure of a
7-layer slab is the same as the substrate of
$\sqrt{2}a\times\sqrt{2}a$ two-dimensional supercell in Fig.
\ref{fig2a}. For TiO$_2$-terminated case, the TiO$_2$ and SrO layers
exchange their vertical positions. Regarding the isolated FeSe
monolayer, its in-plane lattice constants were fixed at the value of
SrTiO$_3$ surface as in Fig. \ref{fig1a}. This corresponds to 3.91
\AA~for the in-plane distance of Se-Se atom. We also checked the
case of 3.8 \AA~as reported in the experiment \cite{xue}. There is
no meaningful change found.

\begin{figure}
\includegraphics[angle=0,scale=0.3]{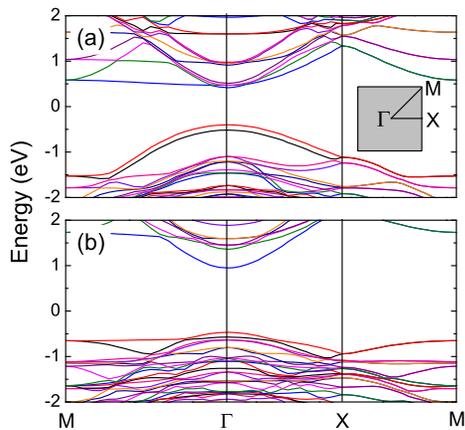}
\caption{(Color online) Band structures of clean
(a)TiO$_2$-terminated and (b)SrO-terminated SrTiO$_3$(001) surface
along high symmetric directions of surface Brillouin zone (SBZ) in
the inset. }\label{fig3a}
\end{figure}

Figure \ref{fig3a} shows the band structures of clean TiO$_2$- and
SrO-terminated SrTiO$_3$ surface in panel (a) and (b), respectively.
The surface Brillouin zone (SBZ) of the $\sqrt{2}a\times\sqrt{2}a$
two-dimensional supercell is shown as the inset of panel (a). In the
following of the paper, we always use the same SBZ since all
structures studied have the same two-dimensional supercell. As can
be seen from panel (a) for TiO$_2$-terminated surface, there are two
nearly degenerated valence bands separated from the conduction bands
by a gap of about 1 eV. These two valence bands are derived from the
2$p$ orbitals of O atoms in the surface TiO$_2$ layers of the
two-sided slab. The bottom of the conduction band is contributed by
the 3$d$ orbitals of Ti atoms in the deep layers. As a contrast, for
SrO-terminated surface in panel (b), the top valence band is formed
by the 2$p$ orbitals of O atoms in the deep layers while the bottom
conduction band originates from the 3$d$ orbitals of subsurface Ti
atoms. The band gaps of clean SrTiO$_3$ surface are thus determined
by the orbitals of O and Ti atoms for both terminations. These
results are consistent with the previous calculations
\cite{borstel1,borstel2}.

\begin{figure}
\includegraphics[angle=0,scale=0.3]{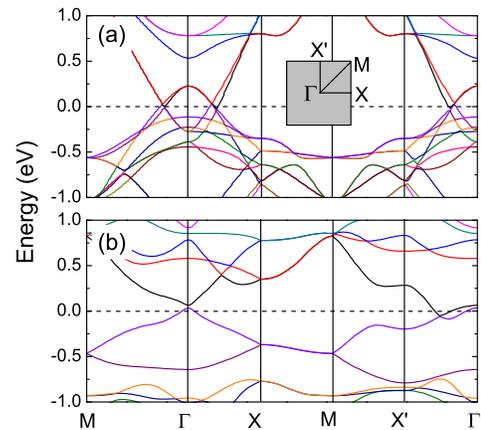}
\caption{(Color online) Band structures of isolated FeSe monolayer
along high symmetric directions of SBZ: (a) nonmagnetic state; (b)
collinear antiferromagnetic state, in which $\Gamma X$ corresponds
to the spins in parallel and $\Gamma X^{\prime}$ to the spins in
anti-parallel. The Fermi energy is set to zero.}\label{fig4a}
\end{figure}

The band structures of isolated FeSe monolayer with lattice
parameter of 3.91 \AA~in the nonmagnetic and collinear
antiferromagnetic states are shown in Fig. \ref{fig4a} (a) and (b),
respectively. As we see, there are two electron-type bands and two
hole-type bands around $\Gamma$ point in the nonmagnetic state,
which is due to the band folding effect since here the unit cell is
$\sqrt{2}\times\sqrt{2}$ unit cell of bulk FeSe on $ab$-plane. In
the collinear antiferromagnetic state, an obvious feature is the
Dirac-Cone-like bands at $\Gamma$ point along $M-\Gamma-X$ with the
Fermi level slightly below the Dirac point and the tiny density of
states at the Fermi level. The isolated FeSe monolayer in the
collinear antiferromagnetic state thus shows a semiconducting
behavior.

\subsection{FeSe on TiO$_2$-terminated SrTiO$_3$}

\subsubsection{Monolayer FeSe}

For monolayer FeSe on TiO$_2$-terminated SrTiO$_3$, the substrate is
simulated with a 6-layer slab as shown in Fig. \ref{fig1a}, which
eliminates one of the two nearly degenerated top valence bands shown
in Fig. \ref{fig3a}(a).

\begin{figure}
\includegraphics[angle=0,scale=0.3]{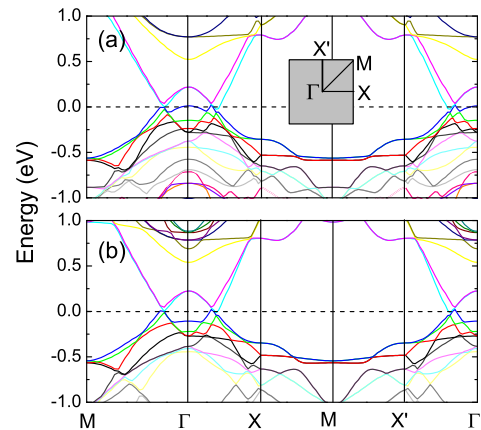}
\caption{(Color online) Band structures of (a)structure I and
(b)structure II for monolayer FeSe on TiO$_2$-terminated SrTiO$_3$
in nonmagnetic state along high symmetric directions of SBZ shown in
panel (a). The Fermi energy is set to zero.}\label{fig0a}
\end{figure}

The band structures of structure I and II (Fig. \ref{fig1a}) for the
monolayer FeSe on TiO$_2$-terminated SrTiO$_3$ in the nonmagnetic
state along high symmetric directions of SBZ are shown in panels (a)
and (b) of Fig. \ref{fig0a}, respectively. For both structures,
there are two electron-type bands and two hole-type bands around
$\Gamma$ point, similar to the case of the isolated monolayer FeSe,
which is clearly due to the band folding effect.
The main difference between (a) and (b) is the relative position of
the band at $\Gamma$ point just below the Fermi level (blue line in
panel (a) and yellow line in panel (b)). This band touches the Fermi
level at $\Gamma$ point for structure I while it falls below the
Fermi level for structure II, which corresponds to the valence band
of the substrate.

By comparing the energies of the nonmagnetic state, ferromagnetic
state, checkerboard antiferromagnetic N\'eel state, and collinear
antiferromagnetic state, we find that the ground state is the
collinear antiferromagnetic state with a large magnetic moment of
2.5 $\mu_B$ on each Fe atom for both structures I and II. In the
ground state, the vertical distances of Se atom to the TiO$_2$ plane
are 3.06 \AA~and 3.16 \AA~in structure I and II, respectively. The
total energy of structure I is lower by ~0.2 eV than that of
structure II. These indicate that the FeSe monolayer has a stronger
bonding with the substrate in structure I than in structure II.

\begin{figure}[!b]
\includegraphics[angle=0,scale=0.3]{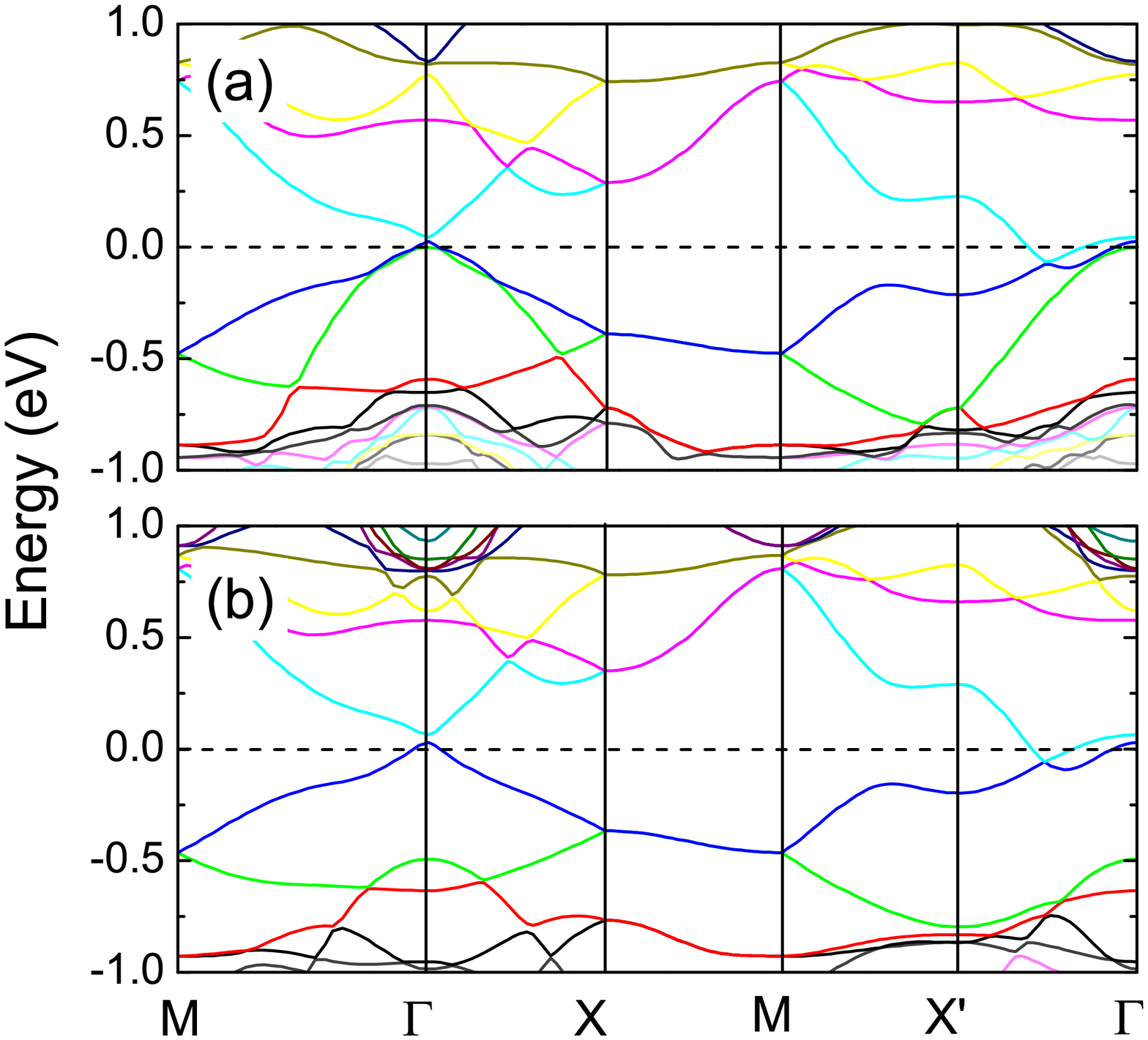}
\caption{(Color online) Band structures of (a)structure I and
(b)structure II for monolayer FeSe on TiO$_2$-terminated SrTiO$_3$
in collinear antiferromagnetic state along high symmetric directions
of SBZ, in which $\Gamma X$ corresponds to the spins in parallel and
$\Gamma X^{\prime}$ to the spins in anti-parallel. The Fermi energy
is set to zero.}\label{fig5a}
\end{figure}

Figure \ref{fig5a} shows the band structures of structure I and II
for the monolayer FeSe on TiO$_2$-terminated SrTiO$_3$ in the
collinear antiferromagnetic state. Compared with Fig. \ref{fig3a}(a)
and Fig. \ref{fig4a}(b), the bands of structure I in Fig.
\ref{fig5a}(a) seems to be a superposition of the bands of separated
substrate and FeSe monolayer. The Fermi level is pinned on the top
of valence band of TiO$_2$-terminated SrTiO$_3$, i.e.
Dirac-Cone-like bands of FeSe lining up the Fermi level with the
parabolic valence band of the substrate. However, such a parabolic
band shifts downwards away from the Fermi level for the structure II
in panel (b).

\begin{figure}
\includegraphics[angle=0,scale=0.3]{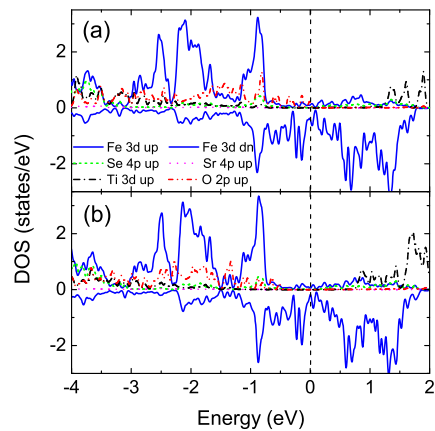}
\caption{(Color online) Orbital-resolved partial density of states
of (a) structure I and (b) structure II for monolayer FeSe on
TiO$_2$-terminated SrTiO$_3$ in collinear antiferromagnetic state.
The Fermi energy is set to zero.}\label{fig6a}
\end{figure}

\begin{figure}
\includegraphics[angle=0,scale=0.22]{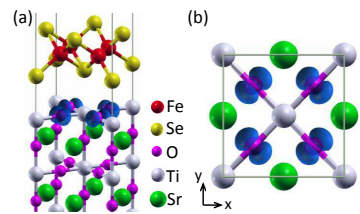}
\caption{(Color online) Side view (a) and top view (b) of band
decomposed charge density isosurface (0.001 e/\AA$^3$) of monolayer
FeSe on TiO$_2$-terminated SrTiO$_3$ in collinear antiferromagnetic
state for the parabolic valence band around $\Gamma$ point in Fig.
\ref{fig5a}(a).} \label{fig7a}
\end{figure}

In order to clarify the contributions to the bands near Fermi level
from different orbitals, we plot the orbital-resolved partial
density of states for structures I and II of the monolayer FeSe on
TiO$_2$-terminated SrTiO$_3$ in the collinear antiferromagnetic
state in panels (a) and (b) of Fig. \ref{fig6a}, respectively. In
both panels, the Fe atom 3$d$ orbitals contribute to the density of
state around the Fermi level, which is located at a valley.
The main difference between panels (a) and (b) is that the 2$p$
orbitals of surface O atoms (red dash dot dot line) in
TiO$_2$-terminated SrTiO$_3$ show states near the Fermi level for
structure I, while they are absent for structure II. This indicates
that the parabolic valence band just below the Fermi level in Fig.
\ref{fig5a} belongs to surface O atoms, which is further confirmed
by the band decomposed charge density shown in Fig. \ref{fig7a}. As
can be seen from the side and top views in Fig. \ref{fig7a}, this
parabolic valence band originates from the $p_x$ and $p_y$ orbitals
of surface O atoms in the TiO$_2$ termination layer.

\subsubsection{Bilayer FeSe}

\begin{figure}[!b]
\includegraphics[angle=0,scale=0.3]{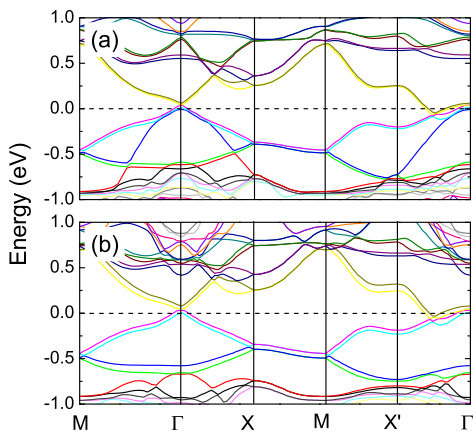}
\caption{(Color online) Band structures of (a)structure I and
(b)structure II for bilayer FeSe on TiO$_2$-terminated SrTiO$_3$ in
collinear antiferromagnetic state along high symmetric directions of
SBZ, in which $\Gamma X$ corresponds to the spins in parallel and
$\Gamma X^{\prime}$ to the spins in anti-parallel. The Fermi energy
is set to zero.}\label{fig8a}
\end{figure}

For bilayer FeSe on TiO$_2$-terminated SrTiO$_3$, the possible
positions of first layer of FeSe are structure I and II in Fig.
\ref{fig1a} while the second layer of FeSe arranges according to
their bulk positions. We still refer them to structure I and II in
the following.

Figure \ref{fig8a} shows the band structures of structure I and II
for bilayer FeSe on TiO$_2$-terminated SrTiO$_3$ in the collinear
antiferromagnetic state. Compared with the band structures of the
monolayer FeSe on TiO$_2$-terminated SrTiO$_3$ in Fig. \ref{fig5a},
the number of bands near the Fermi level contributed by the Fe atoms
is doubled. This feature is similar to the band structure of bilayer
graphene near $K$ point. Moreover, the parabolic valence band near
$\Gamma$ point in panel (a), which originates from the surface O
atoms, shows little shift compared with its corresponding band in
Fig. \ref{fig5a}.

\begin{figure}
\includegraphics[angle=0,scale=0.3]{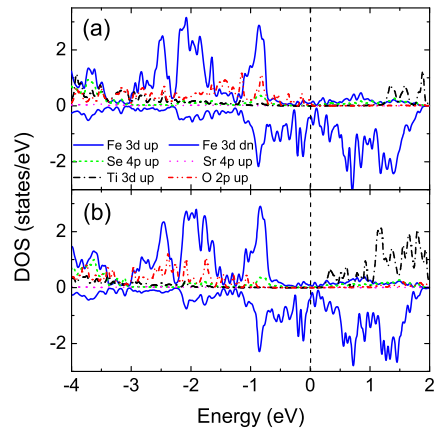}
\caption{(Color online) Orbital-resolved partial Density of states
of (a)structure I and (b)structure II for bilayer FeSe on
TiO$_2$-terminated SrTiO$_3$ in collinear antiferromagnetic state.
The Fermi energy is set to zero.} \label{fig9a}
\end{figure}

The orbital-resolved partial density of states for structure I and
II of the bilayer FeSe on TiO$_2$-terminated SrTiO$_3$ in the
collinear antiferromagnetic state are shown in panels (a) and (b) of
Fig. \ref{fig9a}, respectively. Similar to the monolayer case, the
Fe atoms contribute to the density of state around the Fermi level,
and
the surface O atoms of structure I in TiO$_2$-terminated SrTiO$_3$
show states near the Fermi level, which are due to the parabolic
valence band near $\Gamma$ point in Fig. \ref{fig8a}.

\subsection{FeSe on SrO-terminated SrTiO$_3$}

\subsubsection{Monolayer FeSe}

\begin{figure}
\includegraphics[angle=0,scale=0.3]{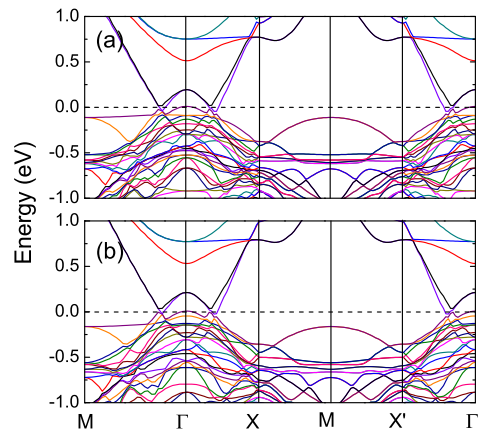}
\caption{(Color online) Band structures of (a)structure I and
(b)structure II for monolayer FeSe on SrO-terminated SrTiO$_3$ in
nonmagnetic state along high symmetric directions of SBZ. The Fermi
energy is set to zero.}\label{fig0b}
\end{figure}

The band structures of structure I and II for the monolayer FeSe on
SrO-terminated SrTiO$_3$ in the nonmagnetic state along high
symmetric directions of SBZ are shown in panels (a) and (b) of Fig.
\ref{fig0b}, respectively. For both structures, there are two
electron-type bands and two hole-type bands around $\Gamma$ point of
the SBZ. The parabolic valence band just below the Fermi level comes
from the valence band of the substrate, the other bands crossing the
Fermi level again due to the band folding effect.


By comparing the energies of the nonmagnetic state, ferromagnetic
state, checkerboard antiferromagnetic N\'eel state, and collinear
antiferromagnetic state, we find that the ground state of monolayer
FeSe on SrO-terminated SrTiO$_3$ is also the collinear
antiferromagnetic state with a large magnetic moment, similar to the
case of TiO$_2$ termination. In this state, the vertical distances
of Se atom to the TiO$_2$ plane are 3.67 \AA~ and 3.64 \AA~ in
structure I and II, respectively. The total energy of structure II
is slightly lower by 0.06 eV than that of structure I.

\begin{figure}[!b]
\includegraphics[angle=0,scale=0.3]{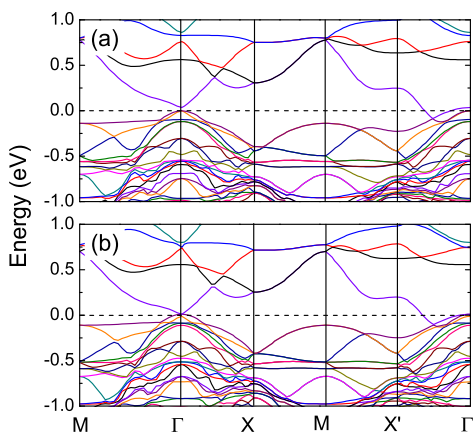}
\caption{(Color online) Band structures of (a)structure I and
(b)structure II for monolayer FeSe on SrO-terminated SrTiO$_3$ in
collinear antiferromagnetic state along high symmetric directions of
SBZ, in which $\Gamma X$ corresponds to the spins in parallel and
$\Gamma X^{\prime}$ to the spins in anti-parallel. The Fermi energy
is set to zero.} \label{fig10a}
\end{figure}

Figure \ref{fig10a} shows the band structures of structure I and II
for the monolayer FeSe on SrO-terminated SrTiO$_3$ in the collinear
antiferromagnetic state. The bands near the Fermi level look like
the superposition of the bands of the substrate in Fig.
\ref{fig3a}(b) and the isolated FeSe monolayer in Fig. \ref{fig4a}(b),
similar to the case of TiO$_2$-termination. However, there is
obvious difference between the bands shown in Fig. \ref{fig10a} and
\ref{fig5a}. This may origin from the different characteristics of
SrO-terminated and TiO$_2$-terminated surfaces, for which Ti-O bond
shows considerable covalency near the surface
\cite{borstel1,borstel2}.

\begin{figure}
\includegraphics[angle=0,scale=0.3]{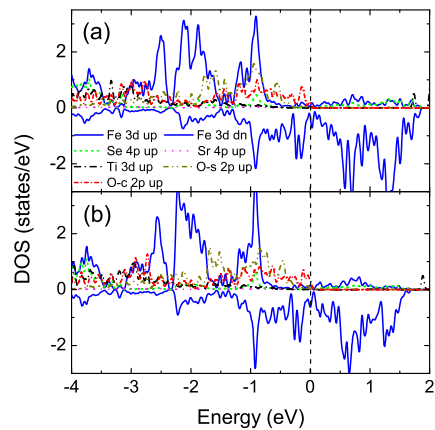}
\caption{(Color online) Orbital-resolved partial density of states
of (a)structure I and (b)structure II for monolayer FeSe on
SrO-terminated SrTiO$_3$ in collinear antiferromagnetic state. The
O-s denotes surface O atom in SrO termination layer and the O-c is O
atom in the TiO$_2$ deep layer. The Fermi energy is set to
zero.}\label{fig11a}
\end{figure}

The orbital-resolved partial density of states for structure I and
II of the monolayer FeSe on SrO-terminated SrTiO$_3$ in the
collinear antiferromagnetic state are shown in panels (a) and (b) of
Fig. \ref{fig11a}, respectively. The O-s denotes surface O atom in
SrO termination layer and the O-c is O atom in the TiO$_2$ deep
layers. In both panels, the Fe atoms contribute most to the density
of state around the Fermi level, which is located at a valley.
However, the O atoms in the TiO$_2$ deep layers contribute more to
the states near Fermi level than the surface O atoms in the  SrO
termination. This is consistent with the previous calculations that
the upper valence band for the SrO-terminated surface is mainly
formed from the orbitals of internal O
atoms\cite{borstel1,borstel2}.

\begin{figure}
\includegraphics[angle=0,scale=0.3]{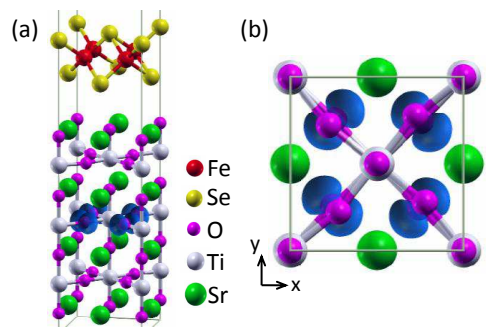}
\caption{(Color online) Side view (a) and top view (b) of band
decomposed charge density isosurface (0.001 e/\AA$^3$) of monolayer
FeSe on SrO-terminated SrTiO$_3$ in collinear antiferromagnetic
state for the parabolic valence band near $\Gamma$ point in Fig.
\ref{fig10a}.} \label{fig12a}
\end{figure}

In order to illustrate the origin of the parabolic valence band near
the Fermi level along $M-\Gamma-X$ direction in Fig. \ref{fig10a},
the band decomposed charge density is plotted in Fig. \ref{fig12a}.
As can be seen from the side view and top view, this parabolic band
originates from the 2$p$ orbitals of O atoms in the TiO$_2$ deep
layers. This is in agreement with the density of states in Fig.
\ref{fig11a}.

\subsubsection{Bilayer FeSe}

Similar to the case of TiO$_2$ termination, for bilayer FeSe on
SrO-terminated SrTiO$_3$, the possible positions of first layer of
FeSe are structure I and II in panels (a) and (b) of Fig.
\ref{fig2a}, respectively. The second layer of FeSe arranges
according to their bulk positions.

\begin{figure}
\includegraphics[angle=0,scale=0.3]{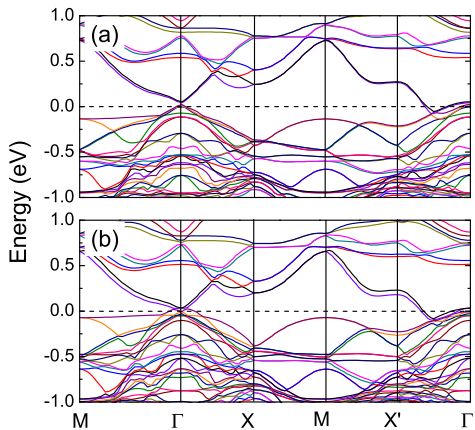}
\caption{(Color online) Band structures of (a)structure I and
(b)structure II for bilayer FeSe on SrO-terminated SrTiO$_3$ in
collinear antiferromagnetic state along high symmetric directions of
SBZ, in which $\Gamma X$ corresponds to the spins in parallel and
$\Gamma X^{\prime}$ to the spins in anti-parallel. The Fermi energy
is set to zero.}\label{fig13a}
\end{figure}

\begin{figure}
\includegraphics[angle=0,scale=0.3]{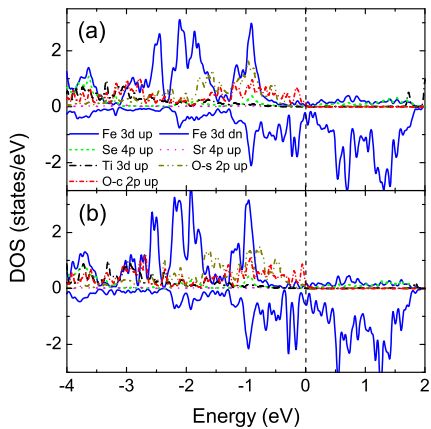}
\caption{(Color online) Orbital-resolved partial density of states
of (a)structure I and (b)structure II for monolayer FeSe on
SrO-terminated SrTiO$_3$ in collinear antiferromagnetic state. The
Fermi energy is set to zero.}\label{fig14a}
\end{figure}

Figure \ref{fig13a} shows the band structures of structure I and II
for the bilayer FeSe on SrO-terminated SrTiO$_3$ in the collinear
antiferromagnetic state. As can be seen near the Fermi level, the
number of bands contributed by the Fe atoms is doubled as well. The
parabolic valence band near $\Gamma$ point in panel (a), which
originates from O atoms in the TiO$_2$ deep layers, shows slight
shift compared with its corresponding band in Fig. \ref{fig10a}.


Figure \ref{fig14a} shows the orbital-resolved partial density of
states for structure I and II of the bilayer FeSe on SrO-terminated
SrTiO$_3$ in the collinear antiferromagnetic state. Similar to the
previous results, the Fe atoms contribute to the density of state
around the Fermi level responsible for the Dirac-Cone-like bands
along $M-\Gamma-X$ direction in Fig. \ref{fig13a}. The 2$p$ orbitals
of O atoms in the TiO$_2$ deep layers also show larger density of
states near the Fermi level than the surface O atoms. Structures I
and II have similar peaks in all energy ranges.


\section{DISCUSSION and Summary}

Inspection of the orbital and layer resolved density of states in
the calculations show that there is no substantial charge transfer
between FeSe layers and the substrate, as expected since both FeSe
layer and SrTiO$_3$ layers are already balanced on the chemical
valences. Thus there is no strong chemical bonding between the FeSe
layer and the substrate SrTiO$_3$. This implies that the electronic
band structure of FeSe ultrathin films on SrTiO$_3$ would result
from the band lineup at the interface between the FeSe thin film and
substrate SrTiO$_3$. The calculations indeed confirm this picture,
as shown in Figs. \ref{fig3a}, \ref{fig4a}, \ref{fig5a},
\ref{fig8a}, \ref{fig10a}, and \ref{fig13a}. The Fermi level of FeSe
thin films is found to locate at the top of the valence band of
substrate SrTiO$_3$. The top valence band of TiO$_2$-terminated
substrate mainly consists of O-$2p$ orbitals within TiO$_2$ surface
layer while O-$2p$ orbitals within deep TiO$_2$ layers for
SrO-terminated substrate. This band can become conducting by a small
amount of hole-doping or by applying a small positive gate voltage.
In the experiment of FeSe-SrTiO$_3$ heterostructure reported in Ref.
\onlinecite{xue}, the superconductivity was found in the FeSe
monolayer absorbed onto the TiO$_2$-terminated substrate, but not in
the bilayer films. Our calculations show that the electronic band
structures of FeSe monolayer and bilayer on TiO$_2$-termination have
similar band structures. Both show a semiconducting behavior. This
may suggest that the observed superconductivity happens either just
in the interface of FeSe-SrTiO$_3$ heterostructures or in the first
layer of FeSe, and the superconducting gap could not be observed in
the double FeSe thin films by the experimental measurements because
the tunneling current was blocked by the second FeSe layer.


\begin{acknowledgments}

We wish to thank Qikun Xue and Dunghai Lee for useful discussions.
This work is supported by National Natural Science Foundation of
China (Grant Nos. 11004243 and 11190024), National Program for Basic
Research of MOST of China (Grant No. 2011CBA00112), the Fundamental
Research Funds for the Central Universities, and the Research Funds
of Renmin University of China (08XNF018, 10XNL016). Computational
resources have been provided by the Physical Laboratory of High
Performance Computing at Renmin University of China. The atomic
structures were prepared with the XCRYSDEN program \cite{kokalj}.

\end{acknowledgments}

\end{document}